\documentstyle[11pt,epsfig]{article}
\columnwidth\textwidth

\begin{document}
\begin{flushright}
FTUV/IFIC-9832\\
hep-ph/9807371
\end{flushright}

\begin{center}
\Large{\bf Neutrino Spin Flavour  Precession in Fluctuating Solar Magnetic Fields.}
\end{center}

\begin{center}
E. Torrente-Lujan. \\
IFIC-Dpto. Fisica Teorica. CSIC-Universitat de Valencia. \\
Dr. Moliner 50, E-461000, Burjassot, Valencia, Spain.\\
e-mail: e.torrente@cern.ch\\
\end{center}

\begin{abstract}

The effect of a 
random magnetic field in the convective zone of the 
Sun on resonant neutrino spin oscillations, i.e.
transitions of the type $\nu_{eL}\to\tilde\nu_{\mu R}$, is considered. 
The average 
survival probability  and 
the expected experimental signals
 in the existing solar neutrino experiments are computed  as a function of the level of the 
noise and magnitude of a constant magnetic field in the convective zone.
From comparison with observed detection rates we conclude that the 
RSFP solutions to the SNP with negligible mixing angle are 
stable under the presence of low or moderate levels of noise. 
Detection rates, specially in the Homestake experiment, are 
however sensitive to large levels of noise. As a consequence, 
an upper limit on 
small scale magnetic fluctuations is obtained
from the combined solar data: 
$\surd\langle  B^2\rangle< 140-200 $ kG for 
the scale $L_0\sim 1000 $ km and 
transition moment $\mu=10^{-11} \mu_B$.

\end{abstract}

\vskip 1cm



\newpage


{\bf 1.} 
A neutrino transition magnetic moment can account, as 
it has been shown in (\cite{akh10}),  for both the observed 
deficiency of the solar neutrino flux and the possible time variations of the signal. 
The overall deficit is caused by the alteration or suppression 
of the neutrino energy 
spectrum.
The time dependence may be caused by time variations of the magnetic 
field in the convective zone of the Sun.

The two main problems for the RSFP scenario are 
the large size of neutrino magnetic moment which is needed and the scarce 
 knowledge which is available   on  
the structure of the magnetic field inside the Sun. 
To overcome the second  problem,
authors usually consider 
different ``plausible'' radial magnetic field  profiles. 
The  absence of a strong poloidal field around the Sun limit
 strongly the maximum value of the magnitude of a magnetic field in its
inner parts.
Therefore, a zero magnetic field is usually considered for the solar core.
In the convective zone, fields of the order of a few kG up to a  
 few tens of kG are usually considered possible. 
 
In this article we will address another aspect 
of the RSFP scenario: the nature 
of the magnetic field. Magnetic fields in nature are very often
 stochastic or
at least strongly chaotic. This happens both, when considering  the magnetic fields coming 
from the seeding of primordial fields or more mundane fields
such as those present in the external shell of the Sun. 
Neutrino propagation in a medium with random magnetic fields has special
 properties, differentiated from the situation where only regular 
magnetic fields are present. The effective Hamiltonian governing the 
evolution of averaged quantities 
become non-hermitic and some sort of dissipation is 
introduced. 
Average conversion probabilities become aperiodic and more sensitive to vacuum parameters in some cases. It has been shown for example that
limits on magnetic moments derived including random magnetic fields in 
Supernova are stronger than those derived using only regular 
fields (see \cite{pas1}). 
On the other hand, for strong noise the situation is in
 some sense opposite, as we will see below in this work the survival probability approaches one half irrespective of the  value of the vacuum parameters.

The aim of this letter  is to study the effect
 of the introduction
of a random magnetic component in a simple  RSFP scenario: 
one which disregards 
neutrino mixing  and assumes that the 
direction of the transverse field does not change along the neutrino 
trajectory.
It has been shown (\cite{akh1}) that the same model with 
a regular magnetic field profile
can provide a rough solution to the SNP. 
The conclusion of \cite{akh1} is that
all the available solar neutrino data can be fitted for 
$\Delta m^2\sim 2 \times 10^{-8}-4\times 10^{-9} $ eV$^2$ with no inner
 field and a convective-zone magnetic field varying 
between $2.5-5 \times 10^4$ G ($\mu=10^{-11}\ \mu_B$).

Summarily, 
the main advantages of models including random magnetic 
fields are connected with: 
a) The no neccesity to
know in detail specific magnetic profiles. b) The potentially stronger effect of the 
random magnetic field in comparison to regular magnetic field: for the latter 
the 
spin precession is roughly proportional to 
the quantity $\exp C\int dr B(r)$ where the integral is 
extended over the region where the 
magnetic field $B$ is present. However for the case of random magnetic fields
the precession is proportional to $\exp C' \int dr B^2(r)$. In addition the 
chaotic precession is irreversible and regeneration effects of the 
original neutrino flux are absent or reduced. 
c) It is still important to assume a large transition 
magnetic moment $\mu$ for the neutrino, but the dependence on it is substantially altered. While for regular oscillations  
the constant  $C\sim\mu$, for random magnetic fields $C'\sim \Omega^2 L_0 \mu^2$. The dependence on the magnetic moment $\mu$ can be traded off, at least partially, for the dependence on the parameters $\Omega^2,L_0$. These parameters describe the randomness and the coherence length of the magnetic field, they can be potentially very large, with an accepted range of variation of $\sim1-3$ orders of magnitude.
d) The 
overall influence of the random nature of the magnetic field is the 
flattening of the neutrino spectrum. 
The strong energy 
dependence on  resonances is reduced, so this scenario is a natural way to avoid strong time variations.

The structure of the article is as follows. First we will write the 
basic neutrino propagation equations in presence of matter and 
magnetic field. In the next section 
a brief description of the corresponding
evolution equation for the averaged density matrix (the
Redfield equation) is given. 
Finally the last 
section is dedicated to show the results of the numerical calculations
and concluding remarks:
The 
Homestake, SAGE-GALLEX (Ga-Ge) 
and (Super)-Kamiokande expected signal  rates 
has been calculated and compared with available experimental 
observations.

\vspace{0.2cm}{\bf 2.} 
  Taking into account theoretical and observational evidence of the 
solar surface  and convection region magnetic fields
it is expected (\cite{park1,park2,vai1,vai2,nic2}) that the 
neutrino, while traveling through the convection zone, will encounter a 
magnetic field of complicated unknown structure 
which will be the sum of large and small (fluctuating) scale components
of sufficiently large amplitude and 
short correlation length.
In this work, we will approximate such a field by the simplest possible 
expression: 
\begin{eqnarray}
B(t)&=&B_0+\tilde{B}(t)
\end{eqnarray}
where $B_0$ is a {\em constant} magnetic field over the full extent of the convective 
zone. $\tilde{B}(t)$ is a fluctuating field with zero average
acting over a relative small layer at the 
bottom of the convective zone
(the unstable transition region between  $r\approx0.65-0.75\ R_0$). 
It is assumed that 
$\tilde{B}(t)$ is a stochastic field: a zero-average 
$\delta$-correlated 
Gaussian process characterized by a two-point correlation function
 parametrized as follows:
\begin{eqnarray} \langle \tilde{B}(t)\tilde{B}(t')\rangle
&=& \langle \tilde{B}^2\rangle L_0\delta(t-t')=\eta B_0^2 L_0 \delta(t-t').
\end{eqnarray}
The length $L_0$ is a basically unknown parameter, values in 
the range $10^3-10^4$ should be allowed.
It has been shown in \cite{tor2} that the $\delta$-correlation function
 is a sufficiently good approximation to more realistic finite correlators
 even for relative large correlation lengths.
The coefficient $\eta$ 
is the ratio between large and small scale fields:  it  
is estimated ( \cite{park1}, p. 517; see also \cite{nic2}) by the expression
\begin{eqnarray}
\eta\simeq \frac{1}{\epsilon} \frac{v}{V_0}
\end{eqnarray}
where $V_0$ corresponds to the Alfven speed in 
the large scale field $B_0$, $v$ 
is the turbulent fluid velocity and $\epsilon$ is a, model dependent, 
undetermined number smaller than 
one. In the turbulent dynamo theory the magnetic field is carried in the 
plasma; equating magnetic and kinetic energy, we obtain $v\approx V_0$, and so 
$\eta> 1$. Some other estimations favor much higher values $\eta>>1$ 
(\cite{semi2}).

In transverse magnetic fields, neutrinos with transition magnetic moments will 
experience spin and flavour rotation simultaneously. For Majorana neutrinos the RSFP converts left-handed 
neutrinos of a given 
flavour into right-handed neutrinos of a different flavour. 
Disregarding flavour 
neutrino mixing by simplicity
the evolution of two Majorana 
neutrinos ($(\nu_{eL}, \overline{ \nu}_{\mu R}$) 
in presence of matter and magnetic fields is described  by the following 
effective Hamiltonian:
\begin{eqnarray}
i\frac{d}{dt}\pmatrix{\nu_{eL} \cr \overline{\nu}_{\mu R}}&=&
\pmatrix{V-\Delta & \mu B_\perp^- \cr \mu B_\perp^+ & 0} \pmatrix{\nu_{eL} \cr 
\overline{\nu}_{\mu R}} 
\label{e1001}
\end{eqnarray}
where
\begin{equation}
 B_{\perp}^\pm(t) \equiv B_x(t) \pm i B_y(t)\equiv |B_{\perp}(t)|e^{\pm i\Phi (t)}
\label{e9001}
\end{equation}
and 
\begin{equation}
\Phi (t) = \arctan B_y/B_x.
\label{e9002} 
\end{equation}
The matter potential is 
 $V=\surd 2 G_F (N_e-N_n)$ where $N_e,N_n$ are the electron and neutron number 
densities, $G_F$ is the Fermi constant and 
$\Delta=\Delta m^2/2E$.
The magnetic field strength $B$ enters the evolution Eq.(\ref{e1001}) being multiplied 
by the neutrino transition moment $\mu$. The existing upper limits on the magnetic moment of the electron neutrino include the laboratory bound
$\mu< 2\times 10^{-10}\mu_B$  from reactor experiments as  well as stronger (one or two orders or magnitude) astrophysical and cosmological limits. 
Expected values of $B\approx 1-100$ kG in the solar convective zone and 
$\mu=10^{-11} \mu_B$ would give an expected range for the 
 product $\mu B\approx 10^{-8}-10^{-6} \ \mu_B G\approx 5.6\ 10^{-7}-10^{-15} $ eV  or in the 
practical units which will be used 
 throughout
 this work $\mu B\approx \ 0.1-10\ \mu_{11} B_4$.

Assuming that the direction of the transverse magnetic 
field does  not change along the neutrino trajectory (absence of twist) the
evolution equation can be written as:
\begin{eqnarray}
i\frac{d}{dt}\pmatrix{\nu_{eL} \cr \overline{\nu}_{\mu R}}&=&
\pmatrix{V-\Delta & \mu B_\perp \cr \mu B_\perp & 0} \pmatrix{\nu_{eL} \cr 
\overline{\nu}_{\mu R}}. 
\label{e1001b}
\end{eqnarray}

In the case of constant magnetic field and matter density, the 
corresponding transition probability obtained from Eq.(\ref{e1001b}) is 
\begin{eqnarray}
P(\nu_{eL}\to\overline{\nu}_{\mu R})&=&\sin^2 (k \mu B_\perp\ t)/k^2
\label{e7001}
\end{eqnarray}
with 
$k^2=1+(\Delta-V)^2/(\mu B_\perp)^2$. In resonance ($\Delta\sim V$) or for 
strong $\mu B_\perp$ the
constant $k\sim 1$ and the probability is a periodic function of $B$ with 
maximum  amplitude $\sim 1$.

The problem described by Eq.(\ref{e1001b}) when 
$V$ is locally varying and
 $B$ is a {\em constant} 
is formally equivalent to a MSW problem with a pseudo mixing 
angle given by a certain function of $\Delta $ and $\mu B$.
In the presence of a matter density of exponential profile, as approximately 
happens in the solar case, and a constant transverse field, the precession 
probability can be obtained in a closed  analytic 
form (see \cite{pet1} for a two 
generation case, \cite{emi1} for a general case). 
In this case the evolution matrix for 
the solution to 
Eq.(\ref{e1001}) is given by 
(detailed expressions can be found in \cite{emi1}):
\begin{eqnarray}
U(t,t_{0})&=& U_s^\dagger(t) U_s(t_0) \label{e1002}, \quad
U_s(t)= \exp (i H_0 t)\ U_r(t) .
\label{e1003}
\end{eqnarray}
The Hamiltonian $H_0$ and effective mixing matrix $u$ are
\begin{eqnarray}
H_0&=& u \pmatrix{\lambda_{+} & 0 \cr 0 & \lambda_{-}} u^\dagger;\quad 
u=  \pmatrix{\cos\theta & -\sin\theta \cr \sin\theta & \cos\theta}
\label{e1003b}
\end{eqnarray}

With
$\lambda_{\pm}=\Delta
 \left (-1\pm\sqrt{1+e^2}\right )/2$; 
$e=2\mu B_\perp/\Delta$,
$\tan^2 \theta=-\lambda_{-}/\lambda_{+}$.
The elements of the matrix $U_r$ are simple 
analytic functions of $\Delta,V$ and the ``mixing angle'' $\theta$.

{\bf 3.} 
The master Equation (\ref{e1001}) 
can be written in terms of the corresponding 2x2 density matrix
$\rho(t)$ as:
\begin{eqnarray}
i\frac{d\rho}{dt}&=&[H_{reg},\rho]+\mu \tilde{B}_x(t)[\sigma_1,\rho]+\mu \tilde{B}_y(t)[\sigma_2,\rho].
\label{e7702}
\end{eqnarray}
The $\tilde{B}_x,\tilde{B}_y$ are the Cartesian transversal components
of the chaotic magnetic field. 
Vacuum mixing terms, matter terms corresponding to the SSM density profile
 and the regular  
magnetic part Hamiltonian, are all included in $H_{reg}$. 
$\sigma_{1,2}$ are the standard Pauli matrices.

We assume that the components 
 $\tilde{B}_x,\tilde{B}_y$ are statistically independent, each of them 
characterized by a $\delta$-correlator.
The averaged evolution equation is a simple generalization (see \cite{tor2} 
for a complete derivation) of the well known Redfield equation
(\cite{lor1})  
for two independent sources of noise 
and reads ( $\Omega^2=\mu^2 \langle \tilde{B}^2\rangle L_0$):
\begin{eqnarray}
i\frac{d\langle \rho\rangle}{dt}&=&[H_{reg},\langle\rho\rangle]-i \Omega^2\left ([\sigma_1,[\sigma_1,\langle\rho\rangle]]+[\sigma_2,[\sigma_2,\langle\rho\rangle]]\right ).
\label{e6115}
\end{eqnarray}

Eq.(\ref{e6115}) can be also written as:
\begin{eqnarray}
i\frac{d\langle\rho\rangle}{dt}&=&[H_{reg},\langle\rho\rangle]-
2i \Omega^2\left 
(2\langle\rho\rangle-\sigma_1\langle\rho\rangle \sigma_1 -\sigma_2\langle\rho\rangle \sigma_2\right ).
\label{e6116}
\end{eqnarray}
In the averaged evolution equations the length $L_0$  appears 
only through the 
product  $\eta L_0$. Therefore,  
we will present our results as a function of the
quantity P which  is a simple function of such a product:
\begin{eqnarray}
P&=&\frac{1}{2}\left (1+ \exp(-\gamma)\right ),\nonumber\\
\gamma&\equiv&4 \Omega^2 \Delta t\equiv 4 \eta L_0 (\mu B_{0})^2/3.
\label{e2003b}
\end{eqnarray}

One reason for using $P$ is that it  is a good approximation for the 
 depolarization that the presence of noise induces in the averaged 
neutrino density matrix. $\Delta t$ is the distance over which the noise is acting.

In general Eq.(\ref{e6116}) must be solved numerically. 
For that purpose we have found advantageous to reduce such equation to a 
single integral equation. The procedure, which  could be of some 
interest for other applications, is outlined in the appendix.

{\bf 4.}  
We have  supposed in our numerical calculation 
that the noise is effective only in a thin layer with thickness 
$\Delta t=0.1\ R_\odot$ starting  at $r=0.65\ R_\odot$.
 Different levels of noise are parametrized by the quantity P (Eq.(\ref{e2003b})) which varies between 1 (no noise) and 1/2 (maximal noise).
This equivalent to a range for 
the r.m.s field 
$\surd\langle \tilde{B}^2\rangle\sim 0-600$ kG
(assuming the
 reasonable value of the  length $L_0=1000$ Km and
$\mu=10^{-11}\mu_B$).

\begin{figure}[p]
\centering\hspace{0.8cm}
\epsfig{file=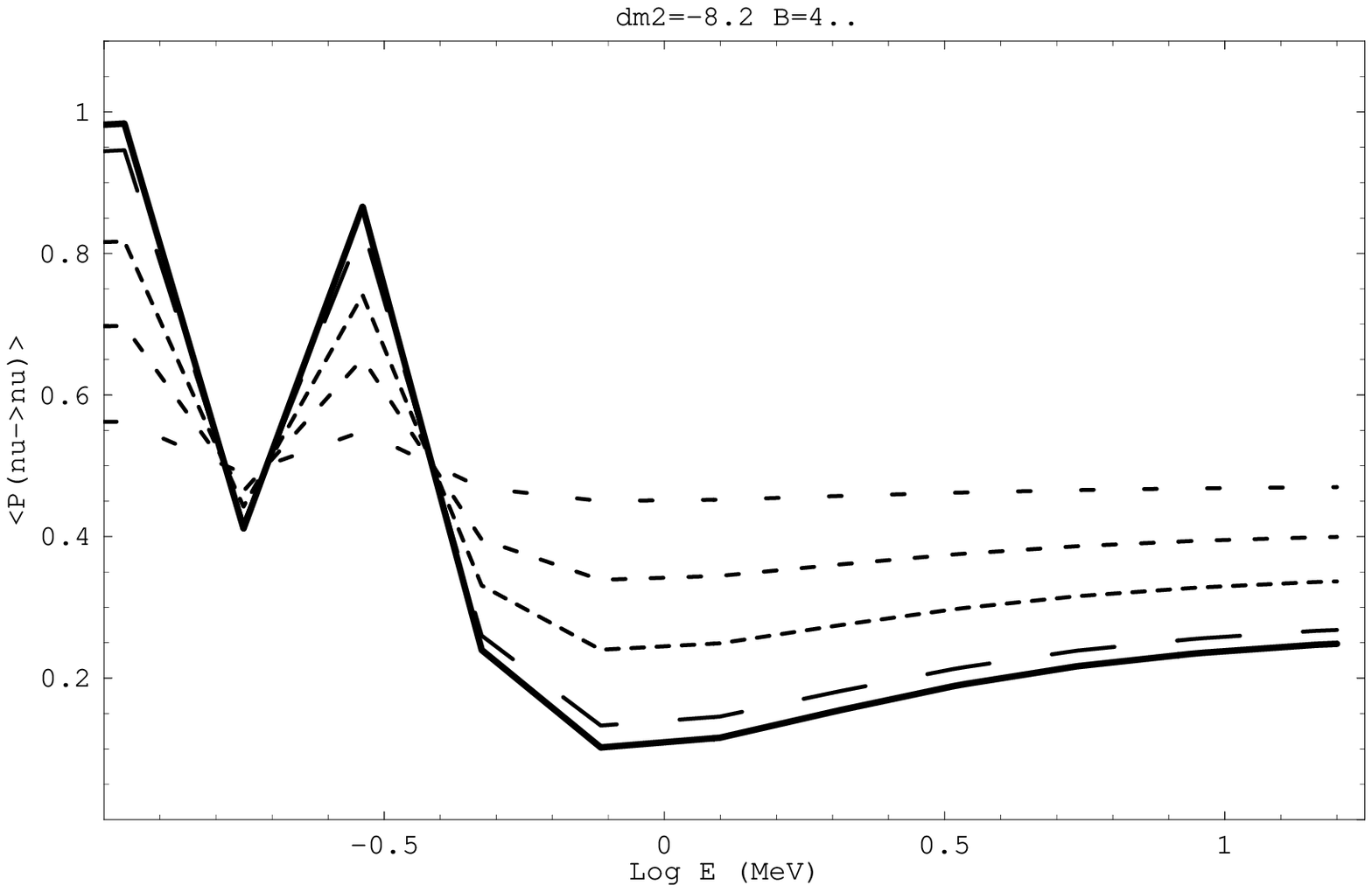,height=12cm}
\caption{
Distortion of the neutrino energy spectrum for different levels of noise:
$P=1.00$ (no noise, solid line); P $=0.95,0.80,0.70,0.50$ (consecutive dashed lines).
The equivalent values for $ \surd\langle \tilde{B}^2 \rangle$ 
are respectively $0.0, 70,150,200,600$ kG (supposing a scale $L_0=1000$ Km,
 $\mu=10^{-11}\mu_B$ and a noise region width $\Delta t=0.1\ R_\odot$).
The corresponding values for the ratio $\eta$ are 
$\eta\equiv\langle B^2\rangle/B_0^2=0,3,14,25,225$. 
It has been used the parameters $\Delta m^2=5.0\ 10^{-9} eV^2$ and
$\mu B_0=4.0\ \mu_{11} B_4$.}
\label{f1}
\end{figure}

Fig.(\ref{f1}) shows
 the distortion of the neutrino 
energy spectrum for different levels of noise. 
We plot the survival 
probability  for an electron neutrino created at 
 the solar interior when arriving at the Earth. 
The neutrino travels in a constant magnetic field 
of magnitude, in this example, $\mu B=4.0\ \mu_{11} B_4$ 
along all the convective zone. An exponential  profile as given by
\cite{BP95} 
is used for the matter density.
This plot has been obtained using the exact solution (Eqs.(\ref{e1003})) for 
the noiseless case and the
numerical solution to  Eq.(\ref{e3001}) for the non-zero noise cases.
 Numerical integration of the original differential equations gives the same results, at a much more higher  CPU cost.

The squared mass difference used in Fig.(\ref{f1})
is $\Delta m^2=5.0\times 10^{-9}$ eV$^2$. This value implies that 
some RSFP occurs at the relevant neutrino energies: For the  
 RSFP  to be   effective in the convective zone it is neccesary that
  $\Delta m^2/2E< 10^{-7}$ eV$^2/$MeV, or for the typical solar 
energies $\Delta m^2< 10^{-8}$ eV$^2$.  
It is clear from  Fig.(\ref{f1}) that the lowest-energy pp-neutrinos with characteristic energy $E_{pp}\approx 0.3 $ MeV, which are detected 
only in the Ga-Ge detectors 
(where they provide more than half of the signal)
 are hardly affected by the magnetic field. 
Intermediate and higher energy neutrinos are strongly suppressed. 
When noise is applied the distortion tends to disappear and  the spectrum 
 becomes practically flat over all the energy range considered
 (for $P<0.7-0.8$). 
One important observation from the figure  is that
the influence of the noisy magnetic field is significant  for all values of 
$\Delta m^2$ even for those whose resonance point is not exactly in the 
region where the chaotic field is present. 

In the next figure, Fig.(\ref{f3}), we show the expected signal 
rates in different neutrino experiments.
The ratio of the averages of the observed detection rates and the SSM expected 
rates which have been used are shown below 
($1\sigma$ statistical and systematic combined errors,\cite{ber1};
the SSM predictions are those of 
\cite{BP95}):
Homestake $ 0.27\pm 0.027$,
SuperKamiokande (345 days)  $0.375^{+0.040}_{-0.002}$,
Ga-Ge (SAGE-GALLEX combined) $0.51\pm 0.06$. 
For computing the expected signal at 
the Kamiokande experiment we have taken into account 
the two elastic scattering processes $\nu_x e\to \nu_x e$ with 
$\nu_x=\nu_e,\overline{\nu}_\mu$.
We have used  the expressions 
for the elastic cross sections 
appearing in \cite{pas2} with a kinetic energy threshold 
for the observation of the
scattered electron of  $T_{e,min}\sim 7$ MeV. 

We observe that the signal is successively correlated 
(situation not favored from the data)
or anti-correlated with 
the value of the magnetic field. 
The periodic structure of the plots  is explained easily if we 
take into account  Eq.(\ref{e7001}).
The variation for the Ga-Ge experiments is always 
significantly lower than those for 
Kamiokande or Homestake as it could be expected from the 
previous discussion about the energy spectrum distortion.
The introduction of noise  smoothes the amplitude of the periodic variations
\footnote{
Of course we could also think of  time variations induced by a 
modulated P.}.

\begin{figure}[p]
\centering\hspace{0.8cm}
\epsfig{file=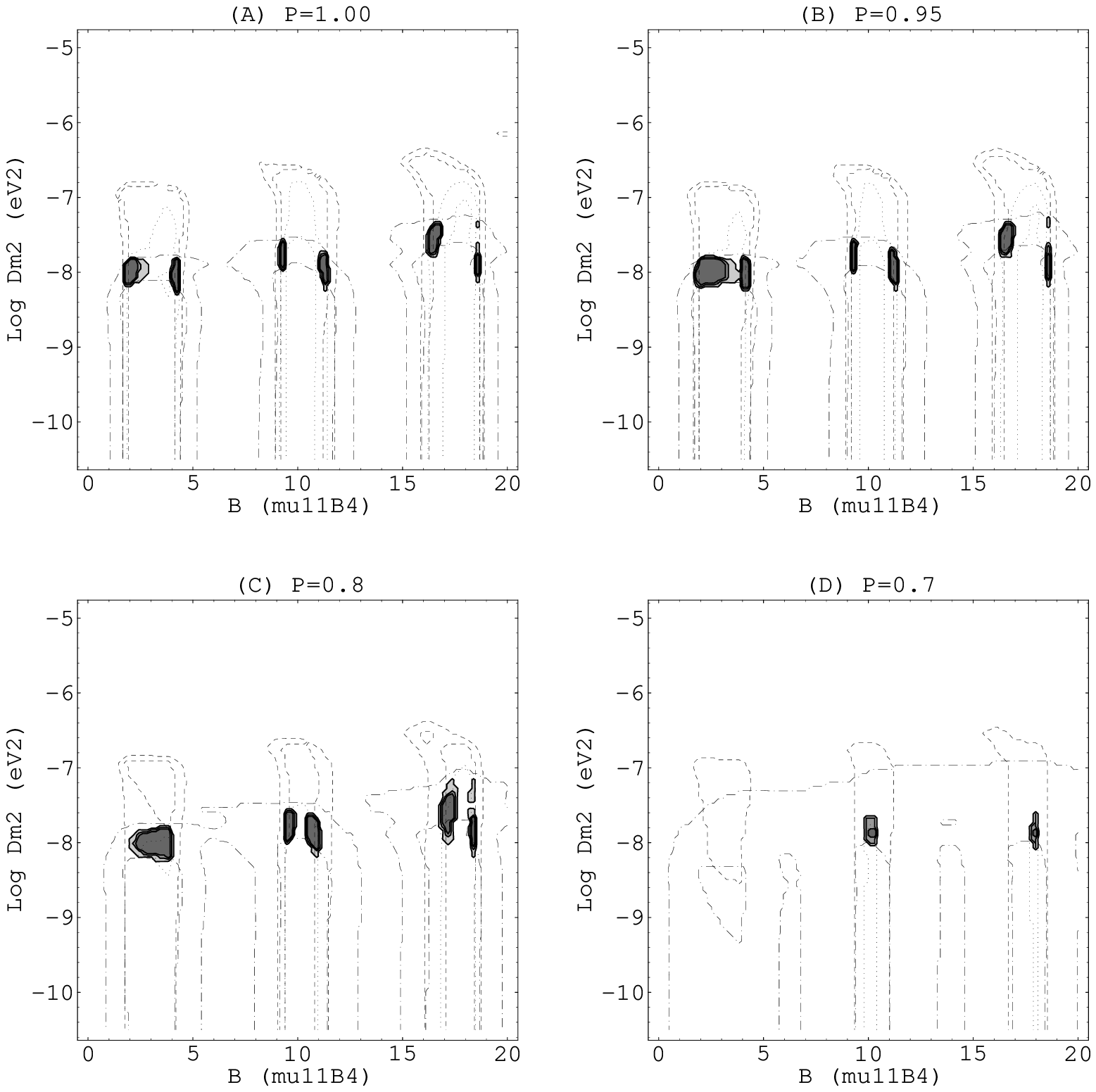,height=16cm}
\caption{ Parameter space regions consistent with the 
observations at the three neutrino experiments. From light to darker shaded 
areas  combined 
$\chi^2$ 90,95,99$\%$ C.L. intervals. 
Lines refer to individual experiments ($95\%$ C.L.): Homestake (dotted line),
 Ga-Ge ( dot-dashed), Kamiokande (dashed).
Plot (A), P=1 (absence of noise). Plots (B,C,D), P=0.95,0.8,0.7.
(See caption in Fig.(\protect\ref{f1})).}
\label{f16}
\end{figure}

In Fig.(\ref{f16}) we present the single and 
 combined $\chi^2$ exclusion plots 
corresponding
to the expected signal rates at the 
three experiments as a function of the oscillation parameter, regular
magnetic field and  fluctuation amplitude.
The Akhmedov  (\cite{akh1}) solutions are reproduced 
 clearly ($\Delta m^2\approx 10^{-8}$ eV$^2$,
 $\mu B\approx 4\ \mu_{11} B_4$ and higher modes).
These solution regions (shaded areas) do not change  with the addition of noise  up to 
a value $P=0.8$. For the value $P=0.7$ we observe 
that the solution with the lower magnetic 
field has disappeared already. For values $P<0.7$ (not shown in the plot)
 the three allowed regions from combined data disappear completely.

If we observe the outcome of the individual experiments (individual lines) 
we can conclude 
that the stronger constraint at high noise comes from the Homestake data.
The other two experiments show contradictory tendencies which can be 
explained by the different weight of the different parts of the neutrino 
spectrum in each of them. While the Kamiokande allowed region tends
to shrink with increasing noise, the region allowed by the Ga-Ge data in fact 
enlarges to  a maximum extension. At $P=0.7$ and low 
magnetic field there is a large common area
which is allowed by both Kamiokande and Ga-Ge data.  

Taking into account the three experiments, and 
within this very simple scenario, particle 
physics solutions to the solar neutrino problem are allowed only if
$P>0.8-0.7$.  
According to the figures, stronger values of the noise are still 
allowed if we admit solutions with ever  larger  magnetic fields.
Large magnetic fields in the convective zone are, if not completely discarded,
 at least disfavored by astrophysics. However 
from the purely neutrino 
experimental 
side  these solutions increase in attractiveness in conjunction with large
 noise  because of the (anti)-correlation smoothening observed in 
 Fig.(\ref{f3}). 

In conclusion we have shown that RSFP solutions to the solar neutrino 
problem with negligible neutrino mixing survive in the presence of 
random magnetic fields of low or moderate level ($P>0.8-0.7$, 
or $\surd \langle B^2\rangle <140-200 $ kG for 
$\mu=10^{-11}\mu_B$ and a fluctuating scale $L_0=1000$ km). 
Conversely, by requiring a particle solution to 
the solar neutrino problem, in the proposed scenario this range of values 
  for P can be interpreted as 
a upper limit 
on the presence  small scale fluctuating magnetic fields in
 the solar convective zone.  
According to  \cite{tor2}  the $\delta$-correlation function
 is a sufficiently good approximation to more realistic finite correlators
 even for relative large correlation length.
Moreover, the quantity $P$ is  connected with the depolarization 
of the density matrix.
It  offers some advantage 
with respect to more ''physical'' parameters 
 as $\Omega^2$ or
$\surd \langle\tilde{B}^2\rangle$: we  expect that  results 
expressed in terms of $P$ are largely independent of the 
concrete modelizing of  the stochastic properties of 
the fluctuating field (in particular the form of the correlator).
Additional, potentially stronger,
 restrictions on the level of noise can be expected when 
considering the experimental bounds in the solar antineutrino flux 
from Kamiokande and the next generation of solar experiments.

\vspace{1cm}

{\bf Acknowledgments}. 

I am grateful to  V.B. Semikoz for many and very useful discussions 
about the nature of the solar magnetic field.
 This work has been supported by DGICYT under Grant 
 PB95-1077 and by  a DGICYT-MEC contract  at Univ. de Valencia. 
Early versions of this work were developed at the Institute Fur 
 Theoretische Physik, Universitat Bern, supported by a grant 
from the Wolferman-Nageli Foundation.

\vspace{1.0cm}
{\Large\bf Appendix}
\vspace{0.2cm}

Our objective in this appendix is to introduce some mathematical 
manipulations in order to obtain an equation adequate for the purpose of 
numerical calculation.
As a preliminary step,
a term can be eliminated from Eq.(\ref{e6116}) if we define a new density 
$\langle\rho_A(t)\rangle$ as:
\begin{eqnarray}
\langle\rho\rangle&\equiv&e^{-4 \Omega^2 (t-t_0) } \langle\rho_A(t)\rangle.
\end{eqnarray}
It is convenient in addition to define 
 a vector representation for the density matrix $\rho$;
we define, using the standard Stokes parameters: 
$$\rho=\  Diag\left( (\rho_{11}-\rho_{22}),\ 2\Re \rho_{12}, \ 2 \Im \rho_{12}\right ).$$
In vectorial form,
 the equation (\ref{e6115}) is:
\begin{eqnarray}
i\frac{d\langle\rho_A\rangle}{dt}&=&
\left (H_{reg}^V -8i \Omega^2 \tau_1\right )\langle\rho_A\rangle,
\label{e6118}
\end{eqnarray}
where $H_{reg}^V$ and $\tau_1$ are  $3\times 3$ matrices. In particular
 $\tau_1$, which is  vectorial 
 equivalent to  $\sigma_1^2+\sigma_2^2$, is of the form
  $\tau_1=i \ Diag(1,0,0)$.

Our strategy will be to solve exactly the problem 
corresponding to the  first term of the Hamiltonian in Eq.(\ref{e6118}) and 
to consider  the second one as a perturbation. We will be able to sum the infinite 
terms of the perturbative series in terms of a 
single unknown-function to be defined later. This unknown function is 
the solution of a certain scalar integral equation. 
We have found this procedure, albeit apparently artificious,  
very advantageous in practical computations.

Let be $U^0(t,t_0)$ the evolution matrix corresponding to the
equation:
\begin{eqnarray}
i\frac{d \langle \rho_A\rangle}{dt}&=& H_{reg}^V\langle\rho_A\rangle
\label{e6121}
\end{eqnarray}
for the solar matter exponential density profile and for $B(t)=B_0$, 
a constant. This 
equation can be solved exactly (see Eqs.(\ref{e1003}-\ref{e1003b})).
Let us consider now the evolution matrix $U$ for the full 
equation Eq.(\ref{e6118}) and its interaction representation counterpart $U_I$ given by:
\begin{eqnarray}
U_I(t,t_0)=U^{0\dagger}(t,t_0) U(t,t_0).
\end{eqnarray}
Our main result, which can be proved easily,  is that the elements of $U_I$ are given by the
following exact formula:
\begin{eqnarray}
U_I(t,t_0)_{ij}&=&\delta_{ij}+(-i\Omega^2)\int_{t_0}^{t}dt_1 
A_{i}(t_1)S_{j}(t_1).
\label{e6123}
\end{eqnarray}
The vector $S$ is the solution of the following integral equation with
kernel $K$:
\begin{eqnarray}
S(t)&=&B(t)+(-i\Omega^2)\int_{t_0}^t dt_1 K(t,t_1) S(t_1).
\label{e6122}
\end{eqnarray}
The vectors $A,B$ are defined as
\begin{eqnarray}
A_i(t)&\equiv& U^{0\dagger}_{i,1}(t,t_0); \quad
B_i(t)\equiv U^{0}_{1,j}(t,t_0); 
\end{eqnarray}
the kernel $K$ is 
\begin{eqnarray}
K(t_1,t_2)&\equiv&\sum_l B_l(t_1) A_l(t_2)
\end{eqnarray}

In symbolic form the solution to  Eq.(\ref{e6122}) can be written as:
\begin{eqnarray}
S&=&\left( 1+ i\Omega^2 K\right )^{-1} B,
\end{eqnarray}
the solution to Eq.(\ref{e6123}) is then, in the same  symbolic form,
\begin{eqnarray}
U_I&=& 1+(-i\Omega^2) \int A\left (1+i\Omega^2 K\right )^{-1} B
\label{e3001}
\end{eqnarray}

This is the concrete equation which has been used as the  basis for 
numerical computations.

\begin{figure}[p]
\centering\hspace{0.8cm}
\epsfig{file=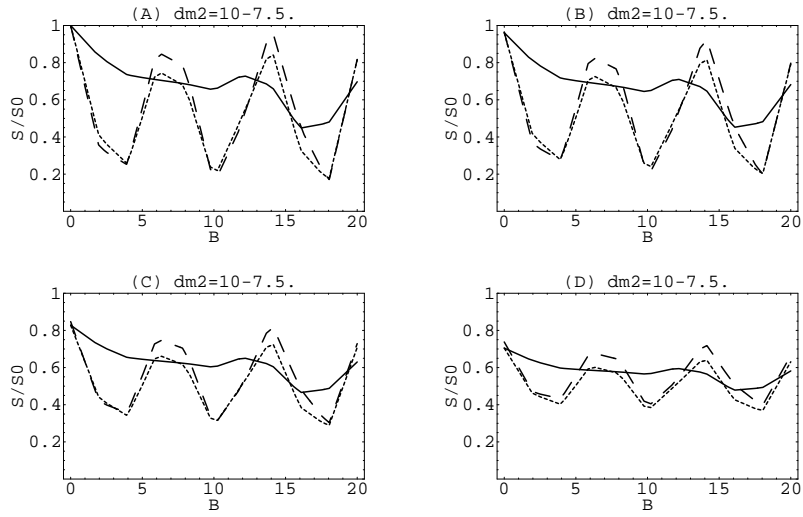,height=16cm}
\caption{The ratio $S/S_0$ (theoretical over observed 
 signal rates) as 
a function of the magnetic field
for the three solar neutrino experiments: Combined Ga-Ge (solid line),
SuperKamiokande (long dashed line) and Homestake (dotted line).
$\Delta m^2=3\times 10^{-8}$ eV$^{2}$. 
Plot(A), P=1 (absence of noise). Plots (B,C,D), P=0.95,0.8,0.7 respectively 
(see caption in Fig.(\protect\ref{f1})).}
\label{f3}
\end{figure}

\newpage

\begin{thebibliography}{10}

\bibitem{akh10}
{ E.K.~Akhmedov}. IC/97/49. 
\newblock {\em Invited talk at the 4th Intl. Solar Neutrino Conference, Heidelberg, Germany,
April 8-11,1997}

\bibitem{pas1}{  S. Pastor, V.B. Semikoz, J.W.F. Valle}.\newblock Physics Letters {\bf B369} (1996) 301-307.
\bibitem{akh1}{ E.K.~Akhmedov, A. Lanza, S.T. Petcov}.
\newblock  Phys. Lett. {\bf B303} (1993), pp.~85--94.

\bibitem{park1}{ E. N. Parker.} \newblock {\em Cosmical Magnetic Fields.} Clarendon Press, Oxford, 1979.

\bibitem{park2}{ E. N. Parker.} \newblock 
Astrophys. J., 408 (1993) 707.

\bibitem{vai1}{ S.I. Vainstein, A.M. Bykov, I.M. Toptygin.} 
\newblock {\em Turbulence, Current Sheets and Schocks in Cosmic Plasma.} 
Gordon and Breach, 1993.

\bibitem{vai2}  S.I. Vainshtein, Y.B. Zeldovich and A.A. Ruzmakin, 
{\em Turbulent dynamo in Astrophysics}. Nauka, Moskow, 1980.


\bibitem{nic2}{ A.~Nicoladis}. \newblock Phys. Lett. {\bf 262} (1991) 2,3, pp.~303-306.

\bibitem{tor2}{ E.~Torrente-Lujan}. \newblock hep-ph/9807361.

\bibitem{semi2}{ E.~Torrente-Lujan, V.B. Semikoz.}. \newblock Work in preparation.

\bibitem{pet1}{  S.T. Petcov}. \newblock Phys. Lett. {\bf B200} (1988) 373..

\bibitem{emi1}{ E.~Torrente Lujan}. 
\newblock Phys. Rev. {\bf D 53}, 4030, (1996).

\bibitem{lor1}{ F.N.~Loreti, A.B. Balantekin}. \newblock Phys. Rev. {\bf D50} (1994), pp.~4762--4770.

\bibitem{BP95} J.N. Bahcall and M.H. Pinsonneault, Rev. Mod. Phys. {\bf 67} (1995) 781.

\bibitem{ber1} V. Berezinsky. astro-ph/9710126.

\bibitem{pas2} S. Pastor, V.B. Semikoz, J.W.F. Valle.
\newblock Phys. Lett. {\bf B423} (1998) 118-125.



\end{thebibliography}
\end{document}